\documentclass[manuscript,authorversion,nonacm]{acmart}
\AtBeginDocument{%
  \providecommand\BibTeX{{%
    \normalfont B\kern-0.5em{\scshape i\kern-0.25em b}\kern-0.8em\TeX}}}

\setcopyright{acmlicensed}
\copyrightyear{2024}
\acmYear{2024}
\acmDOI{XXXXXXX.XXXXXXX}

\acmConference[CHI 2024]{CHI 2024}{May 11--16, 2024}{Honolulu, Hawaii, USA}

\acmISBN{978-1-4503-XXXX-X/18/06}

\begin{document}

\title{MusicTraces: A collaborative music and paint activity for autistic people}
% Music and Paint
% SoundSketch
% Sonic Sketch
% Sonic traces (already taken)
% SonicSensoryMural
% Sonic Palette

\author{Valentin Bauer}
\email{valentin.bauer@polimi.it}
\orcid{0000-0002-3922-7507}
\authornotemark[1]
\affiliation{%
  \institution{Politecnico di Milano, Department of Electronics, Information and Bioengineering}
  \city{Milan}
  \country{Italy}
}

\author{Tommaso Padovano}
\affiliation{%
  \institution{Politecnico di Milano, Department of Electronics, Information and Bioengineering}
  \city{Milan}
  \country{Italy}
}
%\email{larst@affiliation.org}

\author{Mattia Gianotti}
\affiliation{%
  \institution{Politecnico di Milano, Department of Electronics, Information and Bioengineering}
  \city{Milan}
  \country{Italy}
}

\author{Giacomo Caslini}
\affiliation{%
  \institution{Politecnico di Milano, Department of Electronics, Information and Bioengineering}
  \city{Milan}
  \country{Italy}
}

\author{Franca Garzotto}
\affiliation{%
  \institution{Politecnico di Milano, Department of Electronics, Information and Bioengineering}
  \city{Milan}
  \country{Italy}
}

\renewcommand{\shortauthors}{Bauer and Padovano, et al.}

%%Cannot find the number words - aim for 80-120 ?
\begin{abstract} %%here we have 120

Painting and music therapy approaches can help to foster social interaction for autistic people. However, the tools sometimes lack of flexibility and fail to keep people's attention. Unknowns also remain about the effect of combining these approaches. Though, very few studies have investigated how Multisensory Environments (MSEs) could help to address these issues. This paper presents the design of a full-body music and painting activity called ``MusicTraces'' which aims to foster collaboration between people with moderate to severe learning disabilities and complex needs, and in particular autism, within an MSE. The co-design process with caregivers and people neurodevelopmental conditions is detailed, including a workshop, the initial design, remote iterations, and a design critique.
\end{abstract}




\begin{teaserfigure}
  \includegraphics[width=\textwidth]{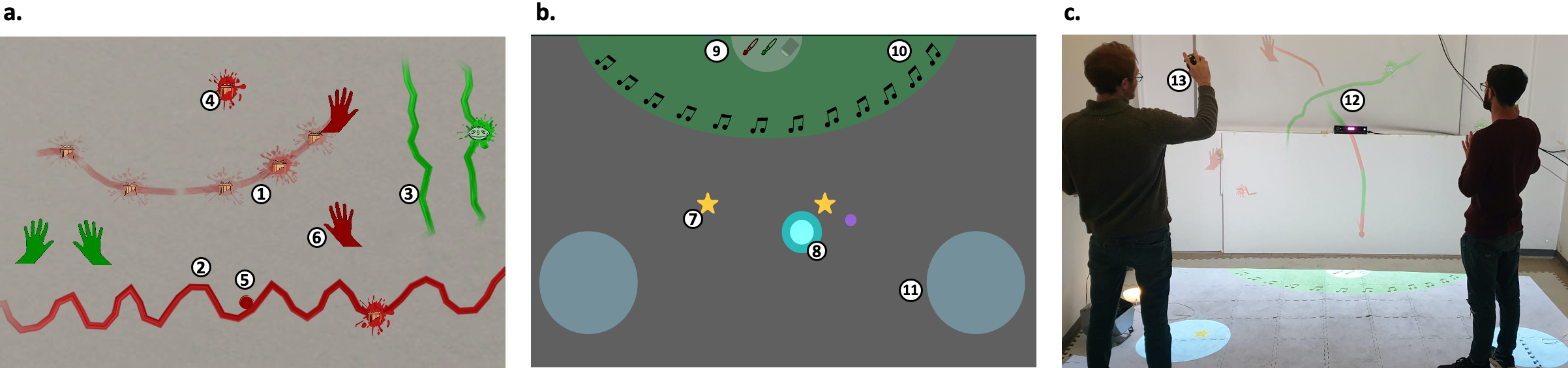}
  \caption{Main design features of the activity MusicTraces. \textbf{(a.)} Front User Interface (UI). 1: Temporary Line. 2: Permanent line (with a sound). 3: Permanent line (without sound). 4: Permanent dot (i.e., sound). 5: Animated dot (i.e., cursor). 6: Hand of red player \textbf{(b.)} Floor UI: 7: player position. 8: Music blob (added after design phase 3). 9: Smart brushes or erasers being detected as held by the users. 10: Floor area to trigger the background music. 11: Floor spots to control the music evolution. \textbf{(c.)} Two users playing together in the room. 12: Line crossing blending their colors. 13. Smart brush.}
  \Description{Application MusicTraces}
  \label{fig:teaser}
\end{teaserfigure}


\maketitle

\section{Introduction} %Skeleton of the section, paragraph by paragraph

%%autism and other conditions
Autism is a Neurodevelopmental Condition (NC) which involves social communication and interaction difficulties and sensory issues \cite{american_psychiatric_association_diagnostic_2013}. Autistic people\footnote{This paper adopts autism stakeholders' language preferences \cite{bottema-beutel_avoiding_2021}, e.g., identity first-language (e.g., autistic people), no offending terms (e.g., ``disorder'').} can display mild learning disabilities and low support needs (e.g., difficulty initiating conversation), or severe learning disabilities and high support needs (e.g., minimal language), sometimes associated with intellectual disability (ID). The latter are little represented in current research \cite{russell_selection_2019}.
%%art practices
Creative practices can help to foster social interaction and self-esteem for people with all kinds of abilities. Painting and music improvised practices are common, and can benefit to people with severe learning disabilities \cite{mayer-benarous_music_2021}. Yet, the tools sometimes lack of flexibility to adapt to people's sensory issues, and unknowns remain about the potential of combining several art practices \cite{bernier_art_2022}.

%%digital tools
Digital tools help to tackle these issues, as being flexible, predictable, and often appealing for autistic people \cite{laurie_international_2019,mazurek_video_2015}. More specifically, Multisensory Smart Environments (MSEs) are promising to promote well-being and social interaction
\citep{Crowell2020Mixed,ringlandSensoryPaint2014,Pares2005Achieving,garzotto2020_magic_room_primary_school,brownCase2016}. They consist of full-body interactive spaces with multisensory stimuli (e.g., visuals, audio, tactile).
% displayed can be precisely controlled, with various devices (e.g., smart objects, projectors, etc.).

%%objectives
This paper presents the design of a full-body music and painting activity called ``MusicTraces'' which aims to foster collaboration within an MSE between young adults with moderate to severe learning disabilities and complex needs, and in particular autism. The MSE consists of a room augmented with floor and front visual projections, speakers, smart objects and lights, and a Microsoft Kinect 2. Below, we report on our co-design process with caregivers and people neurodevelopmental conditions, which includes a workshop, the initial design, remote iterations, and a design critique. 
%followed by research perspectives. 

%To adapt the activity to stakeholder's needs, o
%previously designed with clinical teams
%The space as well as some activities (e.g, association games, storytelling), were previously designed with clinical teams. %More information is available in previous studies \cite{anonymous1,anonymous2}.

%% To create tools close to people's needs, work with stakheolders (Parsons, 2019). Codesign (Frauenberger, 2013).

\section{Related Work}
%%MSE for autism
Several digitally-augmented multisensory settings with full-body interactive capabilities (e.g., smart objects, lights) have been created with clinical teams, with positive outcomes regarding relational aspect for autistic children \cite{Crowell2020Mixed,mora-guiard_sparking_2017,ringlandSensoryPaint2014,Pares2005Achieving,garzotto2020_magic_room_primary_school,cappelen_health_2016}. Some projects, like \textit{Mediate} \cite{Pares2005Achieving}, \textit{Lands of Fogs} \cite{mora-guiard_sparking_2017,Crowell2020Mixed}, or \textit{RHYME} \cite{cappelen_health_2016}, promote multiplayer free exploration of the space. Others like the \textit{Magic Room} \cite{garzotto2020_magic_room_primary_school} or \textit{Sensory Paint} \cite{ringlandSensoryPaint2014} are task-oriented and focus on educative goals.
%, benefitting from serendipitous encounters

%Children can see and act with people around, including practitioners. 
%Creative approaches within MSE
Creative approaches within MSEs have previously been designed for people with NDCs \cite{cappelen_health_2016,ringlandSensoryPaint2014}. \textit{Sensory Paint} is the only project combining music and paint to our knowledge. While focusing more on sonifying the painting experience, it is promising to promote social interaction \cite{ringlandSensoryPaint2014}. \textit{RHYME} consists of various sub projects including musical smart objects (e.g., puppets) affording various interactions (e.g., microphone), with benefits over social aspects and well-being \cite{cappelen_health_2016}. 
%\textit{Sound Forest} is a multisensory music space, with light-emitting interactive strings and vibrating platforms, which seemed to encourage synchronized music-making and collaborative play \cite{frid_sound_2019}. - to use future

%%Project out of MSEs
Other multisensory projects have been designed for autistic children outside of MSEs with similar objectives \cite{cibrian_bendablesound_2017,ragone_evaluating_2022}. \textit{Bendable Sound} allows to play music by touching an elastic display, with benefits over attention and motor development. With \textit{OSMoSIS}, children can play music with their caregiver using their body, to support interactional synchrony \cite{ragone_evaluating_2022}. %\textit{Grapholine} allows people to create musical drawings using one tablet to draw and another to control the drawing parameters (e.g, color), with benefits over  agency.% and educational goals.

Most of the above-mentioned projects consider music as a way of promoting relational aspects within health settings, also called health musicking \cite{stige_notion_2006,stensaeth_musical_2013}. This concept entails multiple ways of experiencing, rather than right or wrong way of playing, closed to the concept of \textit{open work} \cite{stensaeth_umberto_2017,eco_open_1989}. In this context, the system has an active role to support exploration and the co-creative experience (e.g., with hints) \cite{stensaeth_musical_2013}. In particular, the interactive music composition must adapt and evolve based on user actions, such as in the \textit{Reactable} project  \cite{jorda_reactable_2007,villafuerte_acquisition_2012} or \textit{RHYME} \cite{cappelen_djing_2016}. 

At last, to cater for stakeholders' needs, the projects must be co-designed with them \cite{Parsons2020Whose}. This process entails common design principles, e.g., support 
understanding by structuring the information or using video modelling techniques \cite{bellini_meta-analysis_2007}.

%%Mention syntaxes used in previous projects ? Vuzik, Iannix, UPIC

%inspired from  modular systems 
% which uses several music layers (narrative, node, rule).
%Creative approaches for people with disabilities: %%Explain why digital tools can help to complement traditional approaches -> interaction user - system - user (RHYME)
% without digital tools: music, paint
% music approaches: Reactable, Iannix,  -> other Thiebaut
% paint approaches: 
% Other ?
% Music & drawing: Vuzik [Ichino et al., 2014], Grapholine, UPIC [Xenakis, 1977]

%Iannix ? Soundcontrol ? % Music & drawing: Vuzik [Ichino et al., 2014], Grapholine, UPIC [Xenakis, 1977]
% Sound Forest
%%More theoritical: making music together and health musicking
%%Difficulty of being in the middle when creating -> mention of other projects without disabilities
%%%Composing interactive music

%%%add missing information here
\section{Phase 1 - Co-design workshop}
The design process of MusicTraces started with a one-hour workshop organized with five caregivers and three people with neurodevelopmental conditions (NDCs) and Intellectual Disability (ID). It aimed to brainstorm ideas about how to best combine music and painting activities within an MSE to foster collaboration between people with moderate to severe NDC, such as autism. It also intended to validate the consistency of our objective. Three investigators were present: an animator who conducted the session, a secretary who sorted the emerging ideas, and an observer taking notes. After the workshop, the emerging ideas were discussed with an external psychologist with experience in autism.

\subsection{Participants}
Eight participants were recruited (5 males, 3 females): seven from a music association for people with disabilities and one from our network. They include: the head of the association (H), two educators (Ed1 and Ed2), two psychologists (P1, P2), and three people with NDC and ID (PwD1, PwD2, PwD3). PwD1 (23 year-old male) has an Adams-Oliver Disorder inducing ID and lacks of autonomy. PwD2 (44 year-old male) and PwD3 (32 year-old female) both have neurodevelopmental conditions inducing ID, and social and motor issues. The external psychologist (M), who then reviewed the emerging themes, has been working for more than ten years with autistic people at the clinical institute called F\footnote{Anonymized for the submission}. Participants participated voluntarily without being paid after signing consent forms.
%%Detail more! One PwD physically disabled

\subsection{Protocol}
After the participants were introduced to the investigators, they tried some non-creative activities within the MSE of our laboratory (e.g., storytelling) during twenty minutes. Then, they went to a room cleaned from distracting stimuli (e.g., noise), where paint material was added to foster idea generation (i.e., brushes). The animator outlined the workshop's rationale, organization, and rules: (1) generate as many ideas as you can, (2) do not judge ideas, and (3) feel free to express unusual ideas. Participants used ``concept sheets'' to write or draw their ideas, thus accommodating for diverse abilities. If possible, for each idea, they noted the best and worst aspects to obtain more details and promote creativity.

\begin{figure}[h]
  \centering
  \includegraphics[width=.45\linewidth]{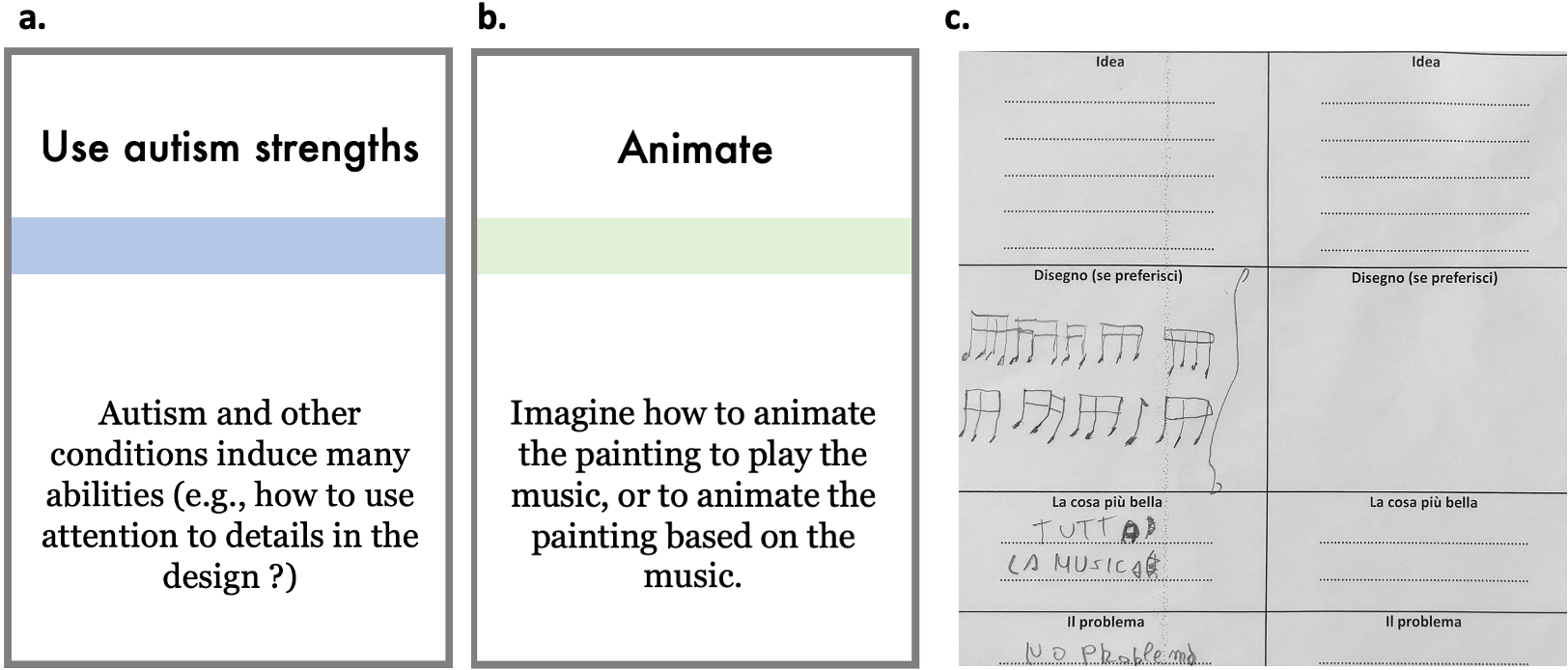}
  \caption{Examples of the material used during the co-design workshop. \textbf{(a.)} Example of activity-related card, \textbf{(b.)} Example of design heuristic card,  \textbf{(c.)} Example of answers on a ``concept sheet'' from a participant with neurodevelopmental condition (PwD1).}
  \Description{Examples of the 20 cards created for the co-design workshop.}
  \label{fig:cards}
\end{figure}

%%Core of the workshop
The brainstorming session involved two 30-minute rounds, first without and then with cards. Both started individually, to note at least one idea without being influenced by others. The participants then shared their idea(s) with others, while also being able to suggest other ideas. The secretary noted the ideas on a Miro board\footnote{Miro application: \url{https://miro.com/fr/}} projected on the wall, and sorted them based on the discussions, inspired from techniques using sticky notes \cite{mackay_chapter_2020}. During the first part, the participants could take inspiration from manipulating the paint material. Then, they could use 20 cards that we previously designed, inspired from previous studies \cite{benton_diversity_2014,scheepmaker_leaving_2021}. Each card bore one design principle and one example of how to apply it. The ten green cards were design heuristics, being picked from an preexisting set \cite{kramer_case-study_2014}. They stated general principles such as ``Allow user to rearrange''. The ten blue cards were design principles related to the activity, inspired from previous studies \cite{scheepmaker_leaving_2021}. For instance, they included 
 ``Think about other music possibilities''. Examples of cards are visible on Figure~\ref{fig:cards}.

%%Ending
The workshop ended with concluding thoughts. The participants were thanked, asked if agreeing on participating in future testing, and given gift bags. Then, they could share a snack with the lab members and try other applications.

%add example of concept sheet filled with the pentagram

%Make the set of cards available

Sessions were filmed. Two authors (secretary and observer) analyzed the data using thematic analysis \cite{braun_using_2006}, with deductive techniques for themes already existing in the literature (e.g., understanding) and inductive techniques to create new ones. Then, they discussed and precised the themes with a third author (the animator) and (M). The identifiers of the themes and the number of participants mentioning them are noted Tx and y/8 (where x and y are numbers).

\subsection{Findings}\label{sec:co_design_findings}
Seven themes were built: \textit{Full-body multisensory interaction} (T1, 5/8), \textit{Music visualization} (T2, 3/8), \textit{Collaboration} (T3, 2/8), \textit{Support} (T4, 2/8), \textit{Gamification} (T5, 3/8), \textit{Understanding} (T6, 3/8), and \textit{Expressivity} (T7, 4/8).

\textit{Full-body multisensory interaction} is expressed by four caregivers and PwD2, to better include people with motor issues. It consists of drawing with the feet (Ed1), or using free (P1) or specific (P2) full-body gestures. Multisensory stimuli are advised (H, P1, P2, PwD2): using tangibles as controllers (e.g., H: a ``stick glowing in the dark''), a microphone (P2), and scents (P1). (M) agreed with this theme, but against using specific gesture, to not confuse people with ID.
%PwD2 added to make clear relationships between the stimuli. 

\textit{Music visualization} is suggested by two caregivers, PwD1, and PwD2. It consists of using a clear visual ``grammar'' for the music that would be appealing and meaningful. For instance, PwD1 suggested to use pentagrams and P2 emphasized to connect every music element to a visual (and conversely). The ExP agreed with this theme.

\textit{Collaboration} is expressed by two caregivers (H, P2), either with turn-taking (H) or playing simultaneously (H, P2). Scenarios could be task-oriented, with users competing (e.g., to learn gesture combinations) (P2, H), or open-ended (e.g., draw how they feel) (H). Though, social anxiety could hinder collaboration (H). (M) agreed with this theme.% remarks, suggesting that turn-taking would not induce too much furstration since the session would be mediated by an adult.
%against each other or time 

% These changes were taken into account, except for the microphone being relegated to future works due to the complexity of directly adding it.
% PwD2 precised the need to establish clear relationships between the different sensory channels
%, for instance to ``recognize materials (such as grass) with sounds'' (T)

\textit{Support} is expressed by two caregivers (H, P2). Based on autism difficulties with abstract thinking, the goal is to prompt creativity by seeing/hearing music or visuals before or during the experience (e.g., painting over a background picture or song). If users get stuck, the caregiver or system should support them. The ExP agrees with this theme.

\textit{Gamification} is evoked by two caregivers and PwD2 to promote engagement. It first consists of having task-oriented use cases (Ed2, P2), such as the discovery of a musical drawing (P2) or a virtual trip (P2). It also concerns the use of a competition logic (P1, PwD2), using levels and challenges (PwD2). ExP agreed with these possibilities. %However, we decided, to not adopt a goal-oriented use case (to not have a right and wrong way to play) but to include some gamiflied design features to promote engagement.

Promoting \textit{understanding} is expressed by two caregivers and PwD2, using smart objects to accommodate for people's abilities (H), or clear rules to lead the music evolution (e.g., based on movements, Ed1). (M) agreed with this theme. He also suggested to map some repetitive movements (e.g., hand flapping) to specific sounds, to give them a meaning.
%PwD2 warned about only using meaningful sounds. 
%PwD2 warned about only using not confusing the users with sounds
%used for reassurance purposes

\textit{Expressivity} is expressed by four caregivers. The goal is to afford symbolization processes, by enabling users to use various movements (P2, Ed1) or painting parameters (e.g., brush colour) based on what they want to convey (P1, H).

At last, the participants stressed before and after the workshop to individualize the design based on tastes (e.g., music) and the objective (e.g., relaxation or communication). ExP advised individualizing the level of multisensory stimuli.

%% add link to the themes from the thematic analysis
\section{Phase 2: Initial design}
MusicTraces is a music and painting activity within an MSE that aims to promote collaboration, inspired from improvised art therapy practices and ``health musicking'' \cite{Stige_health_2012,stensaeth_musical_2013}. As such, it accommodates for multiple ways of being and acting, rather than being task-oriented. The system is considered as a co-creator which supports collaboration with hints \cite{stensaeth_musical_2013}. The design is influenced by the themes from our workshop (noted with [Tx], where x is a number) and previous studies.

\subsection{Environment of the activity}
%%Two main surfaces
This two-user activity is inspired from the contemporary music practice called ``sound painting'' \cite{thompson_soundpainting_2006}. It includes two spaces: the interaction space and the outside space. In the interaction space, delimited with foam carpets [T6], participants can create musicographic objects on a paper-like front user interface (UI) based on their hand position (see Figure~\ref{fig:tablet_brush}.c) [T1]. Smart objects (i.e., brush and eraser) are used to draw [T1], which design was inspired from exiting accessible controllers \cite{ellis_who_2019}. The floor is a control space used to activate the background music [T1,T4]. This design aims to balance stimulation while fostering the engagement of people with severe motor issues (e.g., not moving their arms).
%change the musical instrument or 

The experience relies on multisensory stimuli: audio (from the speakers), visual (with lights and projections), tactile (with the smart objects), and proprioceptive (when drawing and moving) [T1]. Lights become red when users are outside of the interaction space, and turn green when entering it, to prompt agency. All sounds correspond to soothing music instrument based on the literature (i.e., marimba and handpan) to not induce over-arousal \cite{cibrian_step_2018,bauer_music_2023}. All stimuli are simple to avoid cognitive overload. Information is structured in terms of their roles to prompt understanding [T6]. 

The activity is intended for two users, with their caregiver to provide prompts (verbal, physical) and monitor the activity with a tablet (see section~\ref{sec:tablet}), to promote understanding and collaboration [T3, T1]. Some features aim to foster collaboration, e.g., each user has a different color and instrument, the players' trails cross their colors blend [T3].

\subsection{The MusicTraces Syntax}
%%layers
Each visual has a music counterpart and conversely, inspired from [T2] and the \textit{Reactable} project \cite{jorda_reactable_2007}. The experience relies on three layers \cite{andersson_same_2008}: sound node, narrative structure, and composition rules [T6]. Nodes are short music patterns (e.g., notes), narrative structure are combinations of nodes (e.g., melodies), and rules are ways of creating the narratives.

%%design metaphors
Three design metaphors are used [T2, T5, T6]. Short music patterns are represented by paint spots. Melodies are displayed as open or closed lines. They can be played (i.e., as a timeline) when hit by the players. In that case, a cursor - represented by an animated dot - navigates over them at a fixed speed. This metaphor stems from the projects \textit{Iannix} \cite{jacquemin_iannix_2012} and \textit{Upic} \cite{thiebaut_drawing_2008}. Open lines are only heard once when hit, and closed lines multiple times before the cursor fades out. 
%both related to the work of the music composer Xenakis

%%Composition rules
Three composition rules are used [T6]. \textit{Proximity} rules help to combine musicographic element based on their spatial proximity, as in \cite{jorda_reactable_2007}, e.g., circles close to a line are added to the melody. \textit{Harmonization} rules guide the interactive music composition based on users' actions. For instance, users step together on floor circles, which appear after some actions are done [T3], to change the music chord, computed using Markov Chains trained on 37 popular songs\footnote{The dataset can be found at this link: \url{https://www.kaggle.com/datasets/taylorflandro/lyrics-and-chords-from-ultimateguitar?select=pop_lyrics_df.csv}.}. \textit{Rendering} rules take the node position on the ordinate and abscissa axes to affect its pitch (relative to the chord) and positionning.

%%Very simple UI on the floor to explain
\subsection{Interactions}\label{sec:interactions}
MusicTraces allows users to interact in three ways: \textit{Explore}, \textit{Create}, and \textit{Play} [T1,T7]. 
%%explore
\textit{Explore} consists in exploring the music space, by moving the brush and pressing on it. This creates temporary transparent lines that quickly vanish, and triggers sounds based on the brush position.
%%create
\textit{Create} is about leaving permanent paint dots or lines (when using the brush) or erasing them (with the eraser). Creating a spot requires to stay for more than one second in the same position. For the lines, users have to draw a spot while drawing a temporary line (so that every melody contains at least one sound). Users can erase the objects by hovering over them with the eraser.
%%Play
\textit{Play} consists of touching the spots or lines to play the related sounds. When lines are played, a cursor navigates them starting from the interaction point.

\subsection{Hint system}
A \textit{hint system} suggests interactions when it detects \textit{idleness}, \textit{isolation}, or \textit{repetitive movements}, inspired from the \textit{Mediate} project \cite{Pares2005Achieving}, [T3], and [T4]. About \textit{idleness}, three hint levels are used. If a user does no action for more than 20 seconds, the brush slightly vibrates to catch their attention. If they remain idle, it lights up. At last, a hint line appears on the front UI from their hand and with their color. About \textit{isolation}, if they continuously draw in the same area, a notification is sent to the tablet. The caregiver can then choose to display a hint line to redirect their attention elsewhere.
Concerning \textit{repetitive movements}, if a significant similarity is detected (e.g., in terms of shape, length, etc.) between the last line drawn and the others, a notification is sent to the caregiver who can decide whether to display a hint line or not. 
%% the system marks the area as ``used too much''. % for an extended period
%Additionally, to encourage playing music notes, the system identifies line instances where the cursor has not hit any notes for over 5 seconds. In such cases, the cursor gradually fades away as a visual cue. Further discussions on this aspect took place during the next design phases.

% (OLD VERSION) Concerning repetitive movements, two possibilities were imagined to suggest how to exit such behaviours: using
% hint lines to bring attention elsewhere (automatically or controlled by the caregiver, or gradually reducing the effect
% of the movements as in the project called Mediate [ 20 ]. This aspect was discussed further during the next design phases

\subsection{Tablet interface}\label{sec:tablet}
%The tablet is intended for the caregiver to control the activity, stopping things if needed. The UI has three sections. The left panel groups the buttons used to control the game mechanics: pause the game, start/stop the background music, activate floor blobs (appeared before), delete lines, and more. In the middle panel, a representation of the four main areas of the front UI is shown (i.e., top left, top right, etc.). The areas light up in red if overused, or green to suggest the caregiver to trigger a hint line in order to prompt the users to explore this space. To that end, the caregiver can choose between four hint shapes represented below. Finally, the right panel features a notification system. It includes notifications about repetitive behaviours or areas being overused.

\begin{figure}[b!]
  \centering   \includegraphics[width=\linewidth]{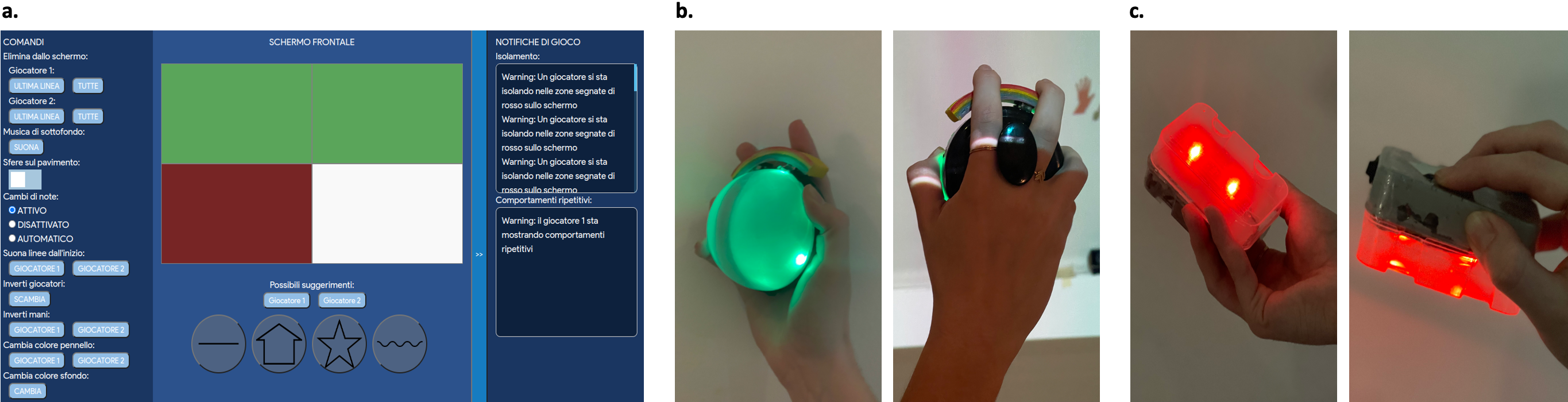}   \caption{Design of the tablet user interface (UI), smart brush, and smart eraser. Elements added after the design phase 4 are mentioned with (D4). \textbf{(a.)} Tablet UI. The Left panel contains buttons to stop the game, control the tutorial steps, remove lines/circles, activate the background music, activate the music evolution (D4), play all melodies, activate the blobs, swap players or hands (security). The center panel manages the hints. The right panel displays the notifications. It can be hidden by clicking on the column at its left (D4). \textbf{(b.)} Smart brush used to draw when pushing on the rainbow button. \textbf{c.} Smart eraser used to erase when holding it.}
  \label{fig:tablet_brush}
  \end{figure}

The caregivers use the tablet UI to monitor the activity or stop it if needed [T4, T6]. The left panel gives control over the game mechanics, e.g., pause the game, remove lines. The middle panel mirrors the four main areas of the front screen, signaling overuse with red lighting or where to trigger hints with green lighting. Four hint shapes are available (e.g., house shape). The right panel is a notification system, alerting about repetitive behaviors or overused areas. The UI was created during the initial design and refined after phase 4. It is visible given on Figure~\ref{fig:tablet_brush}.

\subsection{Apparatus}
%Visual experience (shader, textures) [Valentin]
The activity is developed with Unity engine (version 2021.2.14f1). A custom package gives access to the devices in the MSE by communicating with ad-hoc servers, and to the tablet interface. The web service is developed with Nuxt.js. The tablet UI is built using Vue.js. The activity runs on a windows PC. The hardware components include two projectors, a Microsoft Kinect v2, smart lights, a tablet, and two audio speakers. The four smart objects were built using parts from commercial objects or being 3D printed, magnets, a module ESP8266 Witty Cloud ESP-12F WiFi, external batteries, 2 vibration engines, and LED strips. All sounds were created using virtual instruments on LogicProX Digital Audio Workstation. The paint visuals were created using unity shader graph, with textures made using Inkscape software.

%In particular, Vue.js is here used to build the view layer.%, using HTML, CSS together with JavaScript. 
%3.7V 1800mAh 
% (3V, 80mA, ~11000 ± 2500 rpm)

%a framework to build web applications with Vue.js.
%acts as a proxy giving
% for sending or receiving commands
%module interfaces
%%Most devices operate on an event-based system with subscriptions to events. s
% building user interfaces. Vue.js focuses on the

\section{Phase 3 - Remote agile process}
To be able to conduct future testing in the clinical institute called F, we continued our design process with a psychologist (M) working there, who participated in our previous workshop. The goal was to adapt our activity to this context.

\subsection{Method}
As no similar projects existed to our knowledge, an agile method of working was used with the external psychologist who participated in the first iteration (M) \cite{beck2001agile}. This approach involved doing small iterative design cycles with (M). Email exchanges occurred every two weeks for a total of 3 iterations. Each time, feedback was asked about new changes or ideas stemming from the findings from our workshop and from the related work. To clarify the changes, videos were sent for the two first iterations, segmented by different features. Textual feedback were analyzed using a deductive qualitative analysis process, consisting of analyzing the data according to the activity features \cite{elo_qualitative_2008}. Throughout the three iterations, some features were removed, validated, modified, or added, that are reported below.

%final video link
\subsection{Findings}

%\begin{table}[H]
% \caption{Summary of changes}
%  \label{tab:changes}
%  \begin{tabular}{ p{12em} p{30em} }
%   \toprule
%   \textbf{Category} & \textbf{Details}\\
%   \midrule
%    \textbf{Removed} & (1) Use of other body movements to trigger sounds/visuals. \newline
%      (2) Mimicking the movements of the other player.\\
%    \textbf{Validated} & (1) Overall functioning of the hint system and isolation detection. \newline
%      (2) Adjust music/paint based the distance from the screen. \newline
%      (3) Use of a physical brush to delete lines. \newline
%      (4) Overall dynamics of the game.\\
%    \textbf{Mofidied} & (1) The hint system is changed from "automatic" to "manual from the tablet".\\
%  \textbf{Added} & (1) Closed shapes that get filled with color automatically. \newline
%      (2) Originally, all lines made sounds. Upon proposition to have some lines that don't make any sound, we added them.\\
%  \bottomrule
%\end{tabular}
%\end{table}

%During our 3 iterations of email exchanges, based on the feedback that was given, some features were modified, added or completely removed and some others were kept as approved. Here we report the results of our findings.

Two ideas were abandoned. First, using additional body movements to create sounds and visuals was removed, to not overstimulate the users. Second, triggering specific audiovisual effects when mimicking the other player's was also removed, to not confuse the user since many features were already implemented.

Three element were validated. During the first iteration, the hint system was deemed very useful to help the users. Then, adapting the line thickness based on the user distance from the screen was confirmed, to make the activity more appealing and realistic. The smart objects also received approval and made the caregiver enthusiastic about the project. %Indeed, before having the smart objects, the players had to step out of the interaction space to erase the lines with their hands, which was not ideal for users with severe disabilities. 

The hint system was modified to avoid visual clutter. Indeed, (M) suggested that, after the user draws the first line, the hint lines are no longer automatically added, but rather manually added by the caregiver from the tablet.

Three elements were added to promote agency: closed shapes being automatically filled with color, blobs moving on the floor that make percussive sounds when the users touch them, and lines without nodes. These blobs aim to allow the users being unable to draw with their hands, to be able to make sounds by moving in the space. The lines without nodes are thought for users wanting to create visuals without sounds (as is already possible for sounds).

\section{Phase 4 - Design critique with caregivers}
To further adapt our design to the clinical context of F, and validate its acceptability among a clinical team, a \textit{design critique} was conducted with four caregivers working there: three educators who had not participated in the initial design process and the psychologist included in the third phase (M).% The method is explained below, followed by the findings.

\subsection{Method}
The three educators (Ed3, Ed4, Ed5) have been conducting weekly creative activities at F, for 4 year, more than 10 years, and 5 years. Ed3 (male) and Id4 (male) do group-based manual activities (e.g., painting, sculpting). Ed5 (female) conducts group-based painting activities using a music background, and has a training in art therapy. Id3 specifically works with people with severe disabilities. (M)'s profile was presented above.

%%Protocol
With the educators, individual 45-minute sessions were organized at F's facility, for organizational reasons. The activity features were presented using a video, as well as visuals of the smart object designs and tablet UI (since not fully developed). Sessions started with an activity outline, followed by general questions based on the video steps, and specific inquiries about the hint system and tablet interface. One week later, (M) tested the activity in our lab. After an activity outline, he tested it while making comments, and being asked about his colleagues' ideas.  
%allowing for more insightful feedback compared to emails
% We particularly sought feedback on the tutorial, as we did not know if showing all the game features was the best option. 
Interviews were audio recorded, with caregivers' agreement. The data were then sorted into themes using thematic analysis \cite{braun_using_2006}.

%%remove vibrations smart object
\subsection{Findings}

%%overall feedback
All three educators were positive about using MusicTraces at their facility and participating in the testing. Ed5 said that it closely aligns with her practice and could benefit to her activities. Id3 emphasized the unique combination of music and drawing, noting its potential catch the attention and escape from repetitive behaviors. %Thinking of his clinical experience, Id4 envisioned broader applications for our game (e.g., ) within clinical contexts. 

%%front UI
About the UI, all caregivers agreed with the aesthetic of the front screen. Two suggested changing the background color (Id3 and Id5), as in Id5's painting activities. All of them confirmed to make it possible to change the brush color during the game. All validated the floor UI, regarding the floor blobs and music evolution. To adapt to the users, (M) suggested to make the latter either ``interactable'' (as planned), ``automatic'' (without the floor circles), or disabled.

%Id3 suggested to blend the players' colors where their lines cross. This idea was postponed to perspectives as not deemed directly necessary for our activity (P).
% green background to resemble grass
%At last, Id3 suggested to lower the music level, it was kept since the others were satisfied with it.
%The overall look of the floor's blobs 
% (1) Added a bar that contains information on the instru-

%Interactions
The interactions seemed easy understandable (all), despite some imprecisions in the body tracking (M), that we fixed through coding and with the smart objects. The latter seemed to ease the gameplay (all). (M) noted the colors of environmental lights were too close from the brush colors and prevented from clearly seeing the screen. Thus, we used a different color for the brushes (yellow), and made them brighter outside of the interaction space and darker within.
%leading to potential attention loss. To fix this, 
%Finally, S suggested to blend the colors of the two players when their lines cross at the intersection cross.
%they didn’t really suggest anything about this aspect, but it’s worth reporting that they all agreed the smart objects could really add to the experience. F was really in favor of this, while T said that they were a way for the children to focus their attention (instead of looking elsewhere).

%%hint system
All caregivers validated the use of hints, especially for isolation and repetitiveness. Three aspects were changed. First, hint lines became dashed, to differ from the players' lines. Second, wavy lines were added to guide users' attention toward specific areas, after Ed3's comment. Finally, while all educators enjoyed the use of hints on the smart objects (vibrations, lights), vibrations became an and/off feature, as (M) said that they could confuse or startle some users. %Though, all educators enjoyed the use of slight hints on the smart objects (vibrations, lights). 
%: ``This game does everything to help the caregiver!'' (Id5)
% They all commented on it, saying XXXX. 
%aS in particular suggested also to make a second line appearing in the zone after the “wave” line has disappeared. S did not like the idea of highlighting the selected area in green. 
% and off it was made possible to disable the vibrations from the tablet,
%positively surprised about the features related to isolation or repetitiveness.
% with the educators being especially positive about the features related to isolation or repetitiveness

%%tutorial
About the tutorial, all caregivers agreed on the different steps, the use of video modelling techniques, and the addition of audio recorded instructions. Id3 advised including all steps, contrary to Id4 who said that these ``users cannot keep in mind more than three things.'' Thus, we decided to choose the steps to include before to start with the tablet. (M) agreed with this change. At last, Id3 suggested adding dashed lines that users could follow with their hands.
%since ``some people are not able to replicate only by using the memory'' (Id3), 
%Doing so, the tutorial steps can be gradually introduced over multiple sessions, even if the users can still discover them during playtime. 
%``they cannot keep in mind more than three things at the same time.''

%%tablet interface
All caregivers were positive about the tablet features. The notifications were helpful (Id5). However, the UI contained too much information (M), leading to move some buttons to the left and to allow to hide the notification panel. Indeed, when playing with a user the caregiver would not need to see this panel (M).
%, e.g., to warn about repetitive of behaviours

%%Other insights
One additional insight was suggested by Id3, which (M) agreed on. Since ``some individuals may only stare at the floor'', he suggested to activate only one screen at first, and then complement it with the second screen (e.g., fading in). Though, it was kept for perspectives to not include too many features and potentially confuse the users.

\section{Conclusion and Future Works}
%Summary of what was done
This paper has introduced the design of ``MusicTraces'': an open-ended music and paint activity within an MSE, inspired from improvised art therapy practices, intended to foster collaboration between young adults with moderate to severe neurodevelopmental conditions, and more particularly autism. The co-design process was conducted with caregivers and people with NDC, through a design workshop, the initial design, a remote design process, and a design critique. It also took inspiration from existing studies \cite{jorda_reactable_2007,cibrian_bendablesound_2017}.

%%Contributions
MusicTraces has three main contributions to our knowledge. First, it is the only project that equally combines features about music and paint. Indeed, similar projects leaned more toward paint \cite{ringlandSensoryPaint2014} or music \cite{cappelen_health_2016}. Secondly, it is the only full-body music activity designed for two individuals with NDC, and not a child and their caregiver, as in \textit{Osmosis} \cite{ragone_evaluating_2022}. Thirdly, it is one a the few creative activities intended for people severe conditions, such as \textit{Mediate} \cite{Pares2005Achieving}. Yet, the project is currently bounded to a MSE, inducing issues in terms of portability. Thus, future plans involve porting it to Virtual or Augmented Reality headsets. Including other insights is also considered to support agency, e.g., a microphone control as in \textit{RHYME} \cite{andersson_vocal_2014,cappelen_health_2016}.
%Contribution in the music field that will be explored in future work
%in contrast to most autism research \cite{russell_selection_2019}, 

%Perspectives \\
Next research steps include acceptability and usability testing with people with disabilities, followed by an empirical study with around ten people with moderate to severe NDC at F. A within-group experimental design will be used, where our activity will be compared with a group-based painting activity using background music.

\begin{acks}
The authors would like to thank all the participants who participated in this research. Without them, designing this application would not have been possible.
\end{acks}

\bibliographystyle{ACM-Reference-Format}
\bibliography{sample-authordraft}

% \appendix








% rhoncus. Maecenas eu arcu ac neque placerat aliquam. Nunc pulvinar massa et mattis lacinia.

\end{document}